\documentclass[a4paper,12pt]{article}
\usepackage[utf8]{inputenc}
\usepackage[english]{babel}
\usepackage{authblk}
\usepackage{graphicx}
\usepackage{mathptmx}
\usepackage[singlespacing]{setspace}
\usepackage[headheight=1in,margin=1in]{geometry}
\usepackage{fancyhdr}
\usepackage{url}
\usepackage{makecell}
\usepackage{array}
\usepackage{caption}
\newcolumntype{?}{!{\vrule width 2pt}}
\newlength{\Oldarrayrulewidth}
\newcommand{\Cline}[2]{%
  \noalign{\global\setlength{\Oldarrayrulewidth}{\arrayrulewidth}}%
  \noalign{\global\setlength{\arrayrulewidth}{#1}}\cline{#2}%
  \noalign{\global\setlength{\arrayrulewidth}{\Oldarrayrulewidth}}}
\usepackage{xcolor}

\graphicspath{{images/}}

\title{Cross-platform analysis of user comments in YouTube videos linked on Reddit's conspiracy theory forum}

\author[]{Tomislav Duricic\textsuperscript{1,2}, Volker Seiser\textsuperscript{1}, Elisabeth Lex\textsuperscript{1} \\
\begin{normalsize}\textsuperscript{1}Graz University of Technology, Graz, Austria; \textsuperscript{2}Know-Center GmbH, Graz, Austria \end{normalsize} \\
\begin{normalsize} \{tduricic, elisabeth.lex\}@tugraz.at, v.seiser@student.tugraz.at \end{normalsize}
}

\date{}

\begin{document}

\maketitle
\thispagestyle{fancy}

\begin{center}
\newline
\end{center}

\section*{Extended Abstract}

\noindent \textbf{Introduction and related work.} The rise of social networks and platforms has moved a large part of our social interactions to the online world where we connect with others, engage in discussions, or share opinions and ideas. More often than not, such social interactions are not constrained to a single platform but are rather occurring across several different ones allowing ideas and movements to spread in a cross-platform manner. Even though this free information flow has numerous positive benefits, sometimes it can be malicious in character. For example, a study on Reddit has shown that online communities have the potential of becoming breeding grounds for conflict and anti-social behavior sometimes leading to inter-community conflicts ('wars' or 'raids'), where members of one Reddit community, called "subreddit", collectively mobilize to participate in or attack another community~\cite{kumar2018community}. However, due to many hurdles (e.g., tracking users across multiple platforms, data availability or inconsistency, etc.), there is still a limited understanding of cross-platform information sharing and the influence it has on online discourse. To this end, the authors in~\cite{moyer2015determining} have demonstrated that linking a specific Wikipedia article in a Reddit post is associated with a substantial increase in pageviews. Furthermore, it is worth mentioning that different communities can discuss and frame opinions on the same topics in completely different ways or even have their own lingo\footnote{\label{note1}\url{https://www.reuters.com/article/us-retail-trading-slang-factbox-idUSKBN2AI0JF}}. This phenomenon was demonstrated in a different study on cross-platform communication by hate groups, suggesting that such groups use Facebook for group radicalization and recruitment, while Twitter for reaching a diverse follower base~\cite{phadke2020many}. In this paper, we study the impact that linking YouTube video URLs in Reddit posts (in \texttt{r/conspiracy} subreddit) has on the language used in user comments on those videos. More specifically, we aim to uncover (RQ1) are there noticeable differences in language used before and after the video was posted on Reddit and (RQ2) does the language in user comments on linked YouTube videos become more similar to user language on subreddit where the videos were posted? 

\noindent \textbf{Method.} To create our data set, using the Reddit API, we collect all posts from the \texttt{r/conspiracy} subreddit that reference a URL to a YouTube video uploaded by one of the three channels: \textit{'RussiaToday'}, \textit{'Press For Truth'} and \textit{'corbettreport'}. We do not consider videos with no metadata or that are referenced more than once resulting in a subset of 859 posts and videos. We find that $85\%$ of the videos are linked on Reddit within the first 24 hours after being uploaded on YouTube. Furthermore, using the Reddit API, we collect all user comments in the resulting posts and with YouTube API, all comments posted under the videos are collected. To study changes in the user language, we create three separate text corpora: (i) \texttt{rdt} -- Reddit comments from 859 posts in \texttt{r/conspiracy} (3,008 comments in total), (ii) \texttt{yt\_before} -- comments on 859 YouTube videos that were created before the videos were linked on Reddit (96,452 comments), and (iii) \texttt{yt\_after} -- YouTube comments that were created within the first 24 hours after the videos were linked on Reddit (84,438 comments). To answer RQ1, we compare \texttt{yt\_before} and \texttt{yt\_after} concerning sentiment, subjectivity\footnote{\label{note2}We use TextBlob for sentiment and subjectivity analysis: \url{https://textblob.readthedocs.io/en/dev/}}, two different text corpora representations, and over 200 human-validated lexical categories\footnote{\label{note3}We extract lexical categories from text using Empath: \url{https://github.com/Ejhfast/empath-client}} (topics and emotions). We create two text representations of each text corpus in the following manner. The first representation is created by averaging TF-IDF word distribution after applying tokenization, stemming, and stop words removal to each corpus. The second representation is created by extracting lexical categories, i.e., topical and emotional signals from text using Empath and averaging them over categories for each corpus. To answer RQ2, we compare the distances between text corpora representations using the Jensen-Shannon divergence to see if \texttt{yt\_after} is closer to \texttt{rdt} than \texttt{yt\_before}.

\hspace{-9mm}
\begin{table}[!t]
\begin{minipage}{.5\textwidth}
    \centering\includegraphics[width=\linewidth]{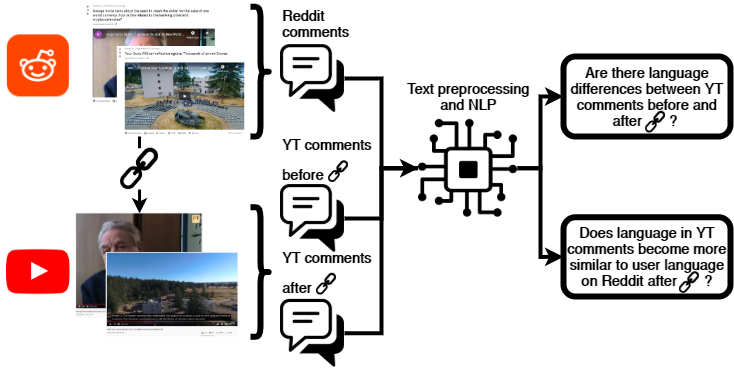}
    \captionof{figure}{Data analysis pipeline -- three user comment text corpora are created from Reddit and YouTube based on their timestamp and the timestamp when the video was posted on Reddit: (i) \texttt{rdt}, (ii) \texttt{yt\_before}, and (iii) \texttt{yt\_after}.}
    \label{fig:methodology}
\end{minipage}
\hspace{4mm}
\begin{minipage}{.47\textwidth}
\resizebox{\columnwidth}{!}{%
    \begin{tabular}{ccc?}
            
            \Cline{1.5pt}{2-3}
            
            \multicolumn{1}{c?}{} & \multicolumn{1}{c|}{\textbf{Sentiment}} & \textbf{Subjectivity} \\ \Xhline{3\arrayrulewidth}  
            
            \multicolumn{1}{?c?}{\texttt{yt\_before}} & \multicolumn{1}{c|}{0.036$\pm$\begin{footnotesize}0.301\end{footnotesize}}            &   0.411$\pm$\begin{footnotesize}0.303\end{footnotesize} \\ \hline
        
            \multicolumn{1}{?c?}{\texttt{yt\_after}} & \multicolumn{1}{c|}{0.036$\pm$\begin{footnotesize}0.297\end{footnotesize}} & 0.413$\pm$ \begin{footnotesize}0.299\end{footnotesize} \\ \Cline{1.5pt}{1-3}
        
            \multicolumn{1}{c?}{\textbf{TF-IDF}} & \multicolumn{2}{c?}{\textbf{Distribution distance}} \\ \Cline{1.5pt}{1-3}
            
            \multicolumn{1}{?c?}{$JSD($\texttt{yt\_before} $||$ \texttt{rdt}$)$} & \multicolumn{2}{c?}{0.2352} \\ \hline
            
            \multicolumn{1}{?c?}{$JSD($\texttt{yt\_after} $||$ \texttt{rdt}$)$} & \multicolumn{2}{c?}{0.2256} \\ \hline
        
            \multicolumn{1}{?c?}{$JSD($\texttt{yt\_before} $||$ \texttt{yt\_after}$)$}  & \multicolumn{2}{c?}{0.0759} \\ \Cline{1.5pt}{1-3}
            
            \multicolumn{1}{c?}{\textbf{Empath lexical categories}} & \multicolumn{2}{c?}{\textbf{Distribution distance}} \\ \Cline{1.5pt}{1-3}
            
            \multicolumn{1}{?c?}{$JSD($\texttt{yt\_before} $||$ \texttt{rdt}$)$} & \multicolumn{2}{c?}{$0.1050$} \\ \hline
            
            \multicolumn{1}{?c?}{$JSD($\texttt{yt\_after} $||$ \texttt{rdt}$)$} & \multicolumn{2}{c?}{$0.0979$}  \\ \hline
            
            \multicolumn{1}{?c?}{$JSD($\texttt{yt\_before} $||$ \texttt{yt\_after}$)$} & \multicolumn{2}{c?}{0.0200} \\ \Xhline{3\arrayrulewidth}  
    \end{tabular}
}
\captionof{table}{Resulting text corpora are compared with respect to sentiment, subjectivity and Jensen-Shannon divergence ($JSD$) between text representations based on TF-IDF and Empath lexical category distributions.}
\label{tab:results}
\end{minipage}
\end{table}

\noindent \textbf{Results and conclusion.} According to the Mann-Whitney U test, there are no significant differences in sentiment and subjectivity value distributions between \texttt{yt\_before} and \texttt{yt\_after}. Furthermore, according to the paired Student's t-test there is a significant difference ($p<0.001$) in TF-IDF representations between \texttt{yt\_before} and \texttt{yt\_after} and $JSD($\texttt{yt\_after} $||$ \texttt{rdt}$)$ $<$ $JSD($\texttt{yt\_before} $||$ \texttt{rdt}$)$. This signals that language used in user comments on YouTube videos has significantly shifted towards language used in \texttt{r/conspiracy} subreddit after posting the video URLs on Reddit. We observe similar results for representations based on lexical categories but without statistical significance. To summarize, in our study, we propose a method for studying content linking from one platform to another and its impact on user language. We show that the language of the first platform slightly changes and becomes more similar to the language where the content was posted. However, our study is not without limitations and in future work, we aim to explore more datasets and different language representation models.

\newpage
\noindent \textbf{Acknowledgements.} This work is supported by the H2020 project TRUSTS (GA: 871481) and the ``DDAI'' COMET Module within the COMET – Competence Centers for Excellent Technologies Programme, funded by the Austrian Federal Ministry for Transport, Innovation and Technology (bmvit), the Austrian Federal Ministry for Digital and Economic Affairs (bmdw), the Austrian Research Promotion Agency (FFG), the province of Styria (SFG) and partners from industry and academia. The COMET Programme is managed by FFG. Preliminary work for this paper was conducted in a team as a mini project during the Summer School on Methods for Computational Social Science organized by GESIS. We would like to acknowledge contribution of all team members, namely Tito Ambyo, Dominik Kowald, Thorsten Luka, and Tomoko Okada. We would also like to thank Mattia Samory for providing the dataset.

\bibliographystyle{ieeetr}

\end{document}